**Grounded world models in biological organisms and future embodied AI**

Giovanni Pezzulo[1,*], Davide Nuzzi[1], Marco D'Alessandro[1], Riccardo Proietti[1], Roberto Bottini[2], Paul Cisek[3]

1. Institute of Cognitive Sciences and Technologies, National Research Council, Rome, Italy
2. Center for Mind and Brain Science (CIMeC), University of Trento, Italy
3. Department of Neuroscience, University of Montreal, Montreal, Quebec, Canada
* Corresponding author: giovanni.pezzulo@istc.cnr.it

## Abstract

Recent advances in generative and embodied AI have been driven by large-scale predictive learning over multimodal data. However, the resulting systems remain largely based on passive training regimes where linguistic regularities create the scaffold onto which information from other modalities is attached. Conversely, neuroscience and cognitive science suggest that biological intelligence is organized in the opposite way, where *grounded world models* acquired through interaction with the environment provide the semantic scaffold to which language is attached. Here, we illustrate five examples of neural circuits supporting grounded world modelling, which underlie navigation in physical and conceptual spaces, affordance-based perception and interaction with objects, active perception and exploratory learning, allostatic control and emotion, and the distinction between self- and world-generated outcomes. These examples highlight several features largely missing from current embodied AI, including the role of intrinsic dynamics as a foundation for learning, the centrality of action in aligning these dynamics with the external world, the prominence of autonomous experience and open-ended learning over passive assimilation of externally provided data, and the fact that early predictive and control mechanisms scaffold higher cognitive abilities such as reasoning, conceptual navigation, planning, imagination, understanding others' minds, and communication. Finally, we discuss whether and how principles derived from biological systems may inform future embodied AI, including training regimes based on social interaction to construct world models that are not only grounded but also socially shared and aligned with human norms and values.

## Introduction

The recent success of large language models has transformed the field of AI, demonstrating the remarkable effectiveness of training large-scale generative models using a predictive learning objective. This approach has enabled the extraction of statistical regularities from human-generated linguistic corpora – regularities concerning the topics humans write and communicate about, including subtle aspects of physical and social reality – in ways that support a wide range of downstream tasks, such as dialogue and reasoning.

This success has motivated efforts to extend the generative AI paradigm to the domain of situated interaction with the physical and social world, commonly referred to as *embodied AI*. It has become increasingly apparent that, when moving to embodied domains, the knowledge available from linguistic corpora alone may be insufficient. Enabling a robot to interpret a sentence such as "pass me the cup, but be careful as it is fragile" and subsequently plan an appropriate course of action may require that AI systems possess *world models* that capture how physical and social reality operates, beyond what can be learned through language alone [1–7].

While the notion of a *world model* encompasses a broad range of concepts – from graphics engines and world simulators to planning systems – we adopt a simplified definition in which its core components are relevant *latent states* (i.e., not directly observable) of the environment and an action-dependent *transition model*. World models capture action–outcome contingencies and enable predictions about how latent states evolve over time, both as a consequence of the agent's actions and through the environment's own dynamics. What world models add to the interpretation of a sentence such as "pass me the cup, but be careful as it is fragile" is knowledge about the structure and dynamics of the physical and social world. This includes spatial representations, such as where the cup is located and whether it can be reached from the current position; object properties, including the persistence of objects over time and their behavior when manipulated; action plans, such as how to grasp and move a cup without breaking it; the communicative intention of the person making the request; and the physical, social, and emotional consequences of damaging a fragile object.

Enabling future embodied AI systems to autonomously acquire world models would not only provide grounding for linguistic instructions, but also unlock a novel set of capabilities beyond those currently available. Such systems could use world models to predict the consequences of their actions and support planning, for example by conditioning on a current state and a desired future state and inferring a likely sequence of actions connecting the two. They could also use world models to generate simulated action–observation trajectories for training controllers in simulation, support reasoning about the environment, and anticipate whether the consequences of one's actions are beneficial or hinder oneself or others – and whether others will appreciate or hate what we do.

One pathway toward developing world models for embodied AI is to scale up the methodology underlying current generative AI systems by integrating multiple modalities within a single generative architecture. This is exemplified by visuo-language-action (VLA) models, which combine visual, linguistic, and action-related information to support planning and goal-directed robotic behavior. This approach is largely based on a predominantly passive training regime, where "passive" refers to the fact that training data are externally selected and curated rather than actively gathered by the learning agent. This phase can be followed by a more interactive stage, in which the agent autonomously engages with the environment and/or with humans, typically over much shorter time spans than the initial pretraining phase. In most current systems, linguistic data play the dominant role in this training pipeline. Indeed, several VLA models are built on top of pretrained large language models, using their representational and generative capabilities as a foundational scaffold. This design choice is motivated by the fact that language models can be trained on extremely large-scale datasets that are not available in other modalities, and have demonstrated strong ability to capture broad regularities that can subsequently support learning in additional modalities. Despite these advances, significant challenges remain, including continual learning, the flexible generalization of knowledge to novel tasks, and robust interaction with humans in real-world environments.

### ***Grounded world models in living organisms***

When considering biological organisms in light of current knowledge in psychology, neuroscience, and related fields, both similarities and differences with present-day generative AI systems become apparent. On the one hand, the predictive learning objective and the construction of generative models bear conceptual resemblance to prominent theories, particularly the framework of predictive processing and active inference [8–10]. In this view, perception, action, and learning can be understood as aspects of a unified inferential process in which biological agents continuously minimize prediction errors either by updating internal models to reflect the world or by interacting with the world to bring it toward a desirable state.

On the other hand, several key ingredients of current generative AI training regimes differ markedly from the processes by which living organisms acquire grounded world models. These include the reliance on largely passive, externally curated learning from an initial tabula rasa state; the fact that autonomous action is typically not integrated into learning until later stages of training; the absence of truly active, continual, or intrinsically motivated learning; and the central role of linguistic data as a scaffold for learning across other modalities. This is exactly the reverse of the phylogenetic and developmental trajectory in humans, where language is grounded in – and builds upon – pre-existing world models acquired through interaction with the physical and social environment.

Living organisms operate under a fundamentally different regime from current generations of generative AI systems. They continuously engage in purposeful action–perception loops to sustain their existence and satisfy their needs and goals. Moreover, their brains and bodies, shaped by long evolutionary processes, impose strong constraints – both enabling and limiting – on learning and cognitive processing. In addition, the world, or more precisely an organism's ecological niche, is inherently dynamic, requiring continual adaptation to address both recurring and novel problems. These problems are rarely identical across instances, and the space of adaptive demands and opportunities continuously expands – especially in species characterized by sophisticated sociality and culture – requiring continual learning, flexible generalization, and robust adaptive capacity. Such flexibility in problem solving has been proposed as a defining feature of intelligence and can be observed across many biological organisms [11].

Living organisms excel in this natural learning regime and are able to develop adaptive strategies that are well suited to their ecological niches. At least part of their evolutionary success can be attributed to the acquisition of *grounded world models*. We use the term "grounded" to emphasize that the knowledge embedded in such models is not merely a distillation of passively obtained sensory inputs or linguistic streams into abstract concepts. Rather, it is grounded in continuous, lived experience realized through open-ended interaction with the environment, which aligns internal brain and bodily dynamics with external states of the world, thereby enabling effective modeling and understanding of that world [2,12,13].

Crucially, embodied cognition theories emphasize that the same grounded world models that support situated prediction and control also provide the foundation for higher cognitive abilities such as problem solving, language processing, advanced mathematical reasoning, and complex social and cultural cognition. For example, neural circuits in the hippocampal–entorhinal system that evolved to support navigation in physical space may have been subsequently redeployed to support memory, planning, and navigation in conceptual spaces [14,15]. Similarly, other cognitive circuits originally evolved for sensorimotor prediction and action control can be temporarily "detached" from the action–perception loop and reused offline to enable higher level cognition. For example, motor systems can be recruited to plan or imagine movements without executing them, or to simulate and infer others' actions and intentions [12,16–19].

This process of abstraction reaches a pinnacle in human language and cognition. In humans, language emerges on top of pre-existing world models acquired through interaction with the environment. While humans can "think with words," such reasoning remains grounded in world models that themselves ground language. It is not only the meaning conveyed by language that is grounded in models of how the world works, but also pragmatic and communicative capacity itself is grounded in models of how the social world works. Linguistic communication builds on a sensorimotor "interaction engine" based on selecting actions that are informative to others, including non-verbal communication and social-cognitive abilities such as intention recognition and turn-taking, many of which precede language [20]. Furthermore, human communication relies on both linguistic and sensorimotor channels, as subtle variations in tone or movement can express intentions, attitudes, or emotions – for example, conveying deference, gratitude, or disappointment through the manner in which we make requests or hand objects to others [21].

In short, in humans language and communication emerge from world models – i.e. exactly the opposite of current embodied AI, where language-derived knowledge is often used as the scaffold for other forms of knowledge and reasoning. To illustrate the fundamental differences between the notions of world models currently used in generative and embodied AI and the grounded world models of living organisms, it is useful to begin with a set of concrete examples and then derive more general lessons.

**Examples of neural circuits supporting grounded world modelling**

The brains of living organisms, from simpler to more complex species, are largely organized to support sensorimotor interactions – namely, identifying action possibilities and predicting, planning, sequencing, and controlling ethological movements.

Not all animals are endowed with sophisticated world models that capture the full complexity of external environmental dynamics. Nevertheless, some form of grounded model appears to be fundamental across organisms. At a minimum, animals require grounded knowledge of their internal milieu to regulate interoceptive variables and maintain homeostasis (see below), as well as knowledge of their own sensorimotor dynamics or of action-outcome contingencies. Yet, such models need not explicitly or faithfully represent the external environment. For example, in C. elegans, whose nervous system comprises fewer than 400 neurons, most neural variability can be explained in terms of a largely preconfigured system for sequencing locomotor behaviors [22] (Figure 1). This can be regarded as an implicit minimal grounded model, in which internal dynamics primarily represent the organism's own locomotor repertoire rather than the richer structure of the external world [23].

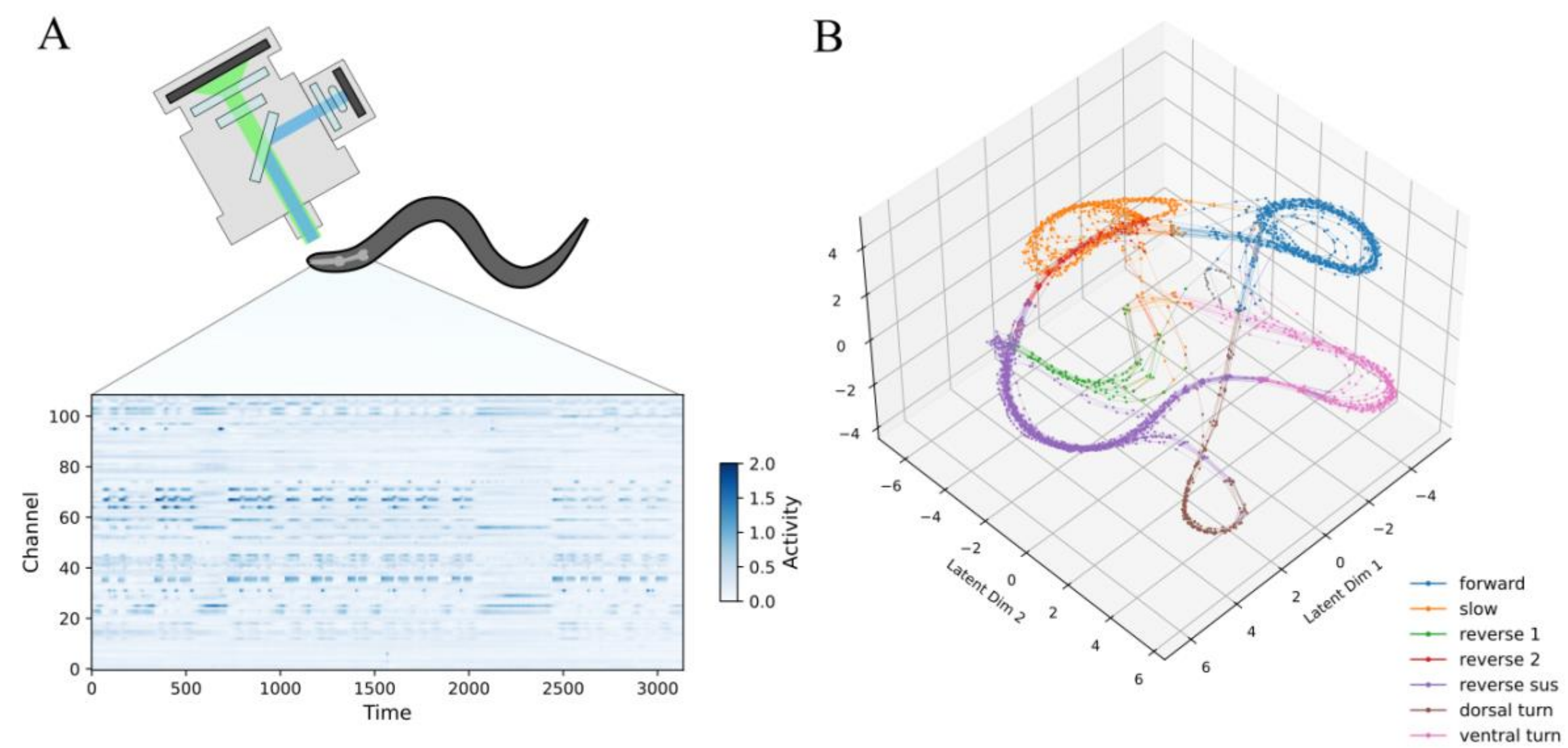

***Figure 1. Whole-brain neural dynamics and a preconfigured behavioral manifold in C. elegans.*** ***A.*** *Schematic of pan-neuronal calcium imaging in a physically immobilized nematode (top), and the corresponding raw activity matrix (bottom) showing normalized fluorescence (ΔF/F) simultaneously recorded from 109 neurons across the head ganglia and anterior ventral cord over the full recording session (data from Kato et al. [22]). Because the animal was restrained within a microfluidic chamber throughout imaging, this activity is generated spontaneously, in the absence of actual movement or environmental feedback.* ***B.*** *Three-dimensional latent manifold obtained by encoding this neural activity with a model trained to (i) generate latent trajectories autonomously and (ii) decode the concurrent behavioral state. Colors denote the seven fictive locomotor states independently annotated by Kato et al. [22] from correlations with kinematics in freely moving animals (forward, slow, reverse 1, reverse 2, reverse sus, dorsal turn, ventral turn); each state occupies a distinct, largely non-overlapping region of the manifold, typically forming closed or near-closed loops. That such a low-dimensional dynamical structure can be recovered from spontaneous activity alone, without any concurrent movement or sensory input, illustrates how the worm's nervous system implements a compact, largely preconfigured model of its own locomotor repertoire.*

In more complex animals, neural organization supports a far larger, more hierarchical, and more diverse repertoire of ethological behaviors, reflecting both increased bodily complexity and the demands of more sophisticated ecological niches. This requires more elaborate circuits for predicting, sequencing, valuating, and selecting among behaviors – and richer grounded world models, which incorporate, for example, knowledge of spatial relations, objects and affordances, as well as other agents and their intentions. From the perspective of predictive processing and active inference, this can be understood as the hierarchical refinement, over phylogenetic and developmental timescales, of increasingly complex generative models for perception and action, built upon more fundamental sensorimotor predictive and control architectures [24]. Unlike current AI systems, this predictive brain architecture is embedded in continuous *spontaneous* brain activity, which it uses for statistical learning and inference. Spontaneous activity – accounting for a large fraction of the brain's energy consumption – is thought to encode statistical priors of the brain's generative model of the world. It shapes perception (predictive coding) and action (active inference) by influencing how stimuli are processed and how actions are prioritized, while also supporting the offline refinement of the brain's models and the progressive alignment of spontaneous dynamics with the natural statistics of the animal's ecological niche, through which organisms become finely attuned to their environments [17,25].

Below, we briefly discuss five examples of neural circuits supporting grounded world modelling, which underlie navigation in physical and conceptual spaces, affordance-based perception and interaction with objects, active perception and exploratory learning, allostatic control and emotion, and distinguishing self- from world-generated stimuli. This is a simplification for illustrative purposes, as cognitive functions are distributed across large-scale interacting brain circuits, but it helps clarify the role of specific components within the broader system.

***Navigation in physical and conceptual spaces***

A foundational idea in cognitive science is that when animals navigate a maze, they do not rely solely on reinforced stimulus–response rules, but instead form a cognitive map – roughly corresponding to a world model of space – that organizes knowledge about the spatial environment and action-dependent transitions, and supports self-localization and the mental construction of navigation plans.

Decades of research have investigated the neural circuits supporting cognitive maps in the hippocampus (e.g., "place cells") and entorhinal cortex (e.g., "grid cells") across rodents, monkeys, humans, and other animals. The grid cell system provides an internally organized reference frame for navigation, autonomously initialized and continuously updated through action, for example via path integration during movement (Figure 2). It implements a coordinate system for spatial navigation on a toroidal manifold [26], enabling a low-dimensional, structured representation of location that supports efficient updating of position and trajectory estimation during movement through space. It operates in coordination with hippocampal place cells, which putatively encode specific locations, jointly forming a cognitive map that supports spatial navigation. The head direction system is another intrinsic system where the organism maintains an estimate of the current head direction based upon the dynamical update of an internal model (typically understood as a 1D manifold or "ring attractor") by acting and subsequently updating the estimate by feedback [27].

The integrated entorhinal–hippocampal system supports multiple aspects of navigation-related world modeling beyond self-localization. Spontaneous sequences of hippocampal activity that "sweep forward" from the animal during navigation ("forward sweeps" [28]) and that depict non-local trajectories during rest ("replay" [29]) have been associated with a range of retrospective and prospective functions, including episodic memory encoding and recall, planning of goal-directed trajectories, and consolidation of both the hippocampal cognitive map and cortical semantic memory.

Interestingly, the integrated entorhinal–hippocampal system appears to be intrinsically organized: its connectivity and dynamics are largely preconfigured and then calibrated and aligned with the external world through action, as suggested by the fact that both grid cell activity and spontaneous hippocampal dynamics can precede navigational experience [17,30,31]. Together, these systems exemplify a mode of learning in which intrinsic neural dynamics are progressively aligned with the structure of the world through embodied action – a process that differs fundamentally from current approaches to training generative AI systems.

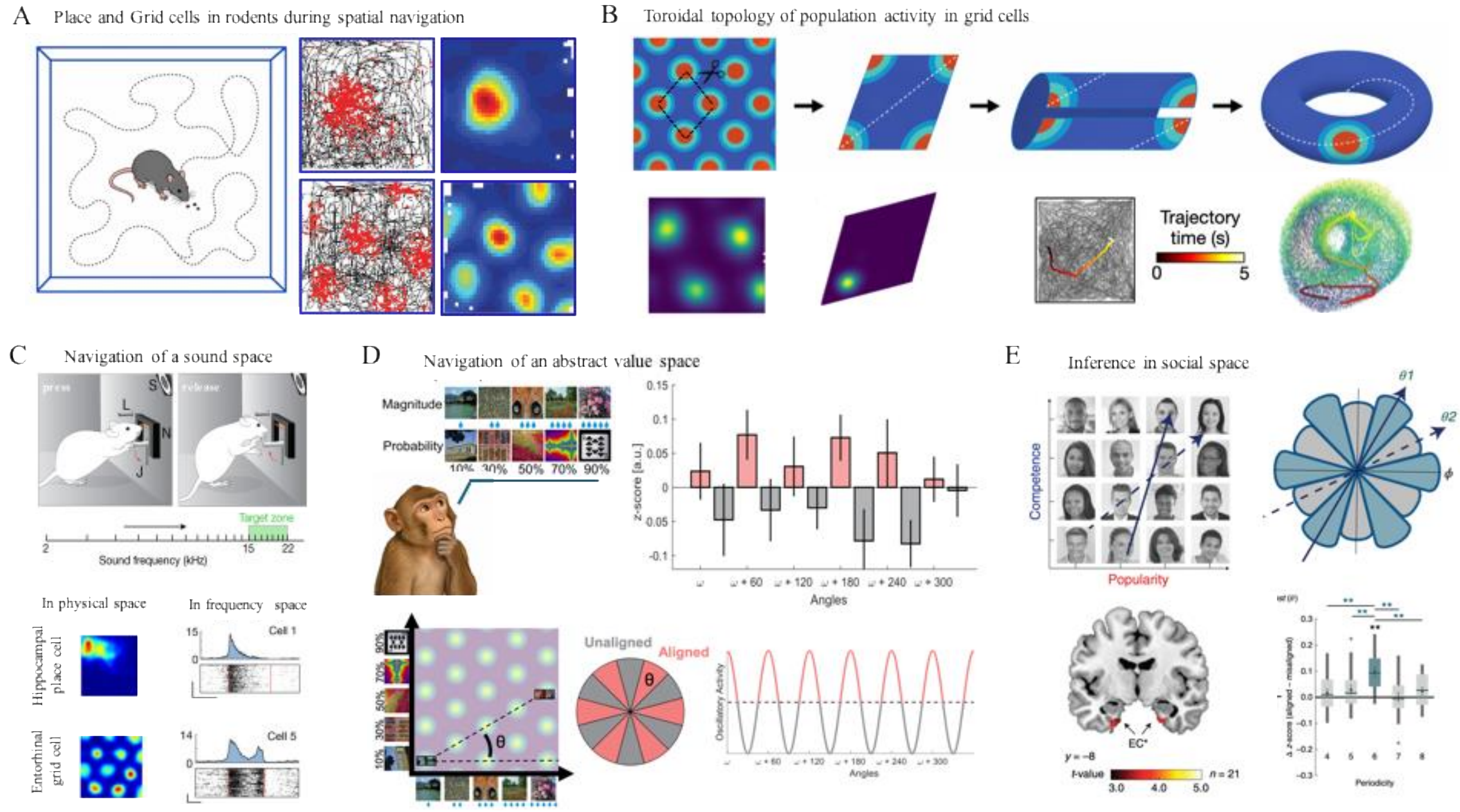

***Figure 2. Spatial and conceptual navigation. A***. *Place-cell and grid-cell firing fields recorded while a rat forages in a regular enclosure. Place cells in the hippocampus typically show localized firing fields, whereas grid cells in the entorhinal cortex fire at multiple locations arranged in a periodic hexagonal lattice. (Figure adapted with permission from [32].)* ***B***. *Toroidal topology of population activity in grid cells. Top row: the periodic lattice of a grid cell defines a fundamental rhombus, whose opposite edges can be identified because they correspond to equivalent phases of the grid pattern. Identifying one pair of opposite edges transforms the rhombus into a cylinder; identifying the second pair transforms it into a torus. In this representation, the firing phase of each grid cell in a module corresponds to a position on the torus, and movement through physical space corresponds to a trajectory that wraps around the toroidal state space (Figure adapted from [33]). Bottom row: examples of grid-cell activity tuned to toroidal coordinates and reconstruction of an animal's trajectory on a torus-like manifold inferred from nonlinear dimensionality reduction of population activity within a grid-cell module (Figure adapted from [26]).* ***C***. *Navigation in a non-spatial sensory space. Rats learn to press a lever to initiate a sound whose pitch changes over time and to release the lever when a target frequency is reached. Hippocampal and entorhinal neurons that show place-like or grid-like firing during spatial navigation can also be preferentially activated at single or multiple locations along the sound-frequency axis (Figure adapted from [34]).* ***D***. *Navigation in an abstract value space. A macaque chooses between pairs of images associated with different reward magnitudes and reward probabilities. The task can be represented geometrically as movement through a two-dimensional value space, in which trajectories can be aligned or misaligned with the axes of a hexagonal grid. Neural activity shows a six-fold modulation as a function of trajectory angle, consistent with a grid-like code (Figure adapted from [35]).* ***E***. *Inference in social space. Human participants learn associations between fictional characters and different levels of competence and popularity, and later use this knowledge to make social inferences. The task can be represented as navigation through a two-dimensional social space. During inference, fMRI activity in the entorhinal cortex shows a six-fold modulation compatible with a grid-like code (Figure adapted from [36]).*

Crucially, this same system appears to provide a scaffold for navigation in more abstract spaces [37–39]. In rodents, hippocampal–entorhinal activity can encode positions along non-spatial, task-relevant continua, such as elapsed time or sound frequency, suggesting that the circuit may implement a more general mechanism for organizing experience along continuous dimensions [34,40]. In non-human primates, grid-like geometries have been reported during the construction of abstract value spaces, for example when options are defined by reward magnitude and reward probability [35,41]. In humans, grid-like codes in the entorhinal cortex and associated navigation networks have been observed when participants learn and navigate conceptual spaces defined by continuous experiential dimensions, including visual and acoustic features [42], odor concentration [43], social attributes [36], and other abstract task variables. These findings suggest that the entorhinal–hippocampal system may support not only self-localization in the external world, but also the organization of knowledge into structured state spaces through which agents can move mentally.

Interestingly, it has been proposed that the same action–perception loop that enables continuous map updating during spatial path integration may also operate during non-spatial model-based inference [15], via simulated actions, eye movements or attentional shifts through conceptual space [44–46]. On this view, conceptual thought can be understood as a form of internalized navigation: a partially detached reuse of sensorimotor mapping mechanisms to explore, update, and act upon structured knowledge. More broadly, many forms of "navigation"—through physical space, conceptual space, and even autobiographical memories and imagined future events (prospection)—may exploit conserved neural mechanisms, with the hippocampal–entorhinal system playing a central organizing role.

***Affordance-based perception and interaction with objects***

In living organisms, a central function of perception is not only to construct a description of the external world, but to identify action possibilities – *affordances* – that capture the tight coupling between perception and action [47]. Affordances can be understood as elements of a predictive model of action possibilities grounded in the perception–action loop. As such, they do not model the world per se, but rather the currently available possibilities for interaction between the agent and the world, capturing how environmental structure relates to the organism's capabilities for action. Indeed, affordances are inherently relational, depending both on environmental structure and on the organism's bodily properties, such as size, strength, and skill. An object that is reachable, graspable, or climbable for one animal may be inaccessible or non-actionable for another.

At any point in time, an animal is surrounded by many affordances – places to which one could run, objects one could grasp, etc. A key task of behavior is to select from among these one that, when exploited, will improve one's state. This kind of "embodied" decision-making can be described as a competition between affordances – a continuous valuation and selection among affordances based on their predicted outcomes [48,49]. This includes both immediate rewards (grasping a fruit and bringing it to the mouth satisfies hunger) as well as new affordances which will be made available (moving toward a fruit to bring it within reach).

At the neural level, the specification of affordances is strongly associated with the dorsal visual stream, long implicated in visuomotor control, while the cues used for selecting which affordances to exploit involve the ventral visual stream, implicated in object recognition, and frontal cortical regions[48] (Figure 3). This functional division of labor is found across mammals [50–52], and may be a basic distinction in all vertebrates [53]. Interestingly, recent studies suggest that in the dorsal stream there exist partially distinct circuits for species-typical actions (e.g. for primates: circuits for feeding, reaching, climbing, object manipulation, etc.) [54,55] reminiscent of the behavior-based architectures of autonomous robotics[56,57].

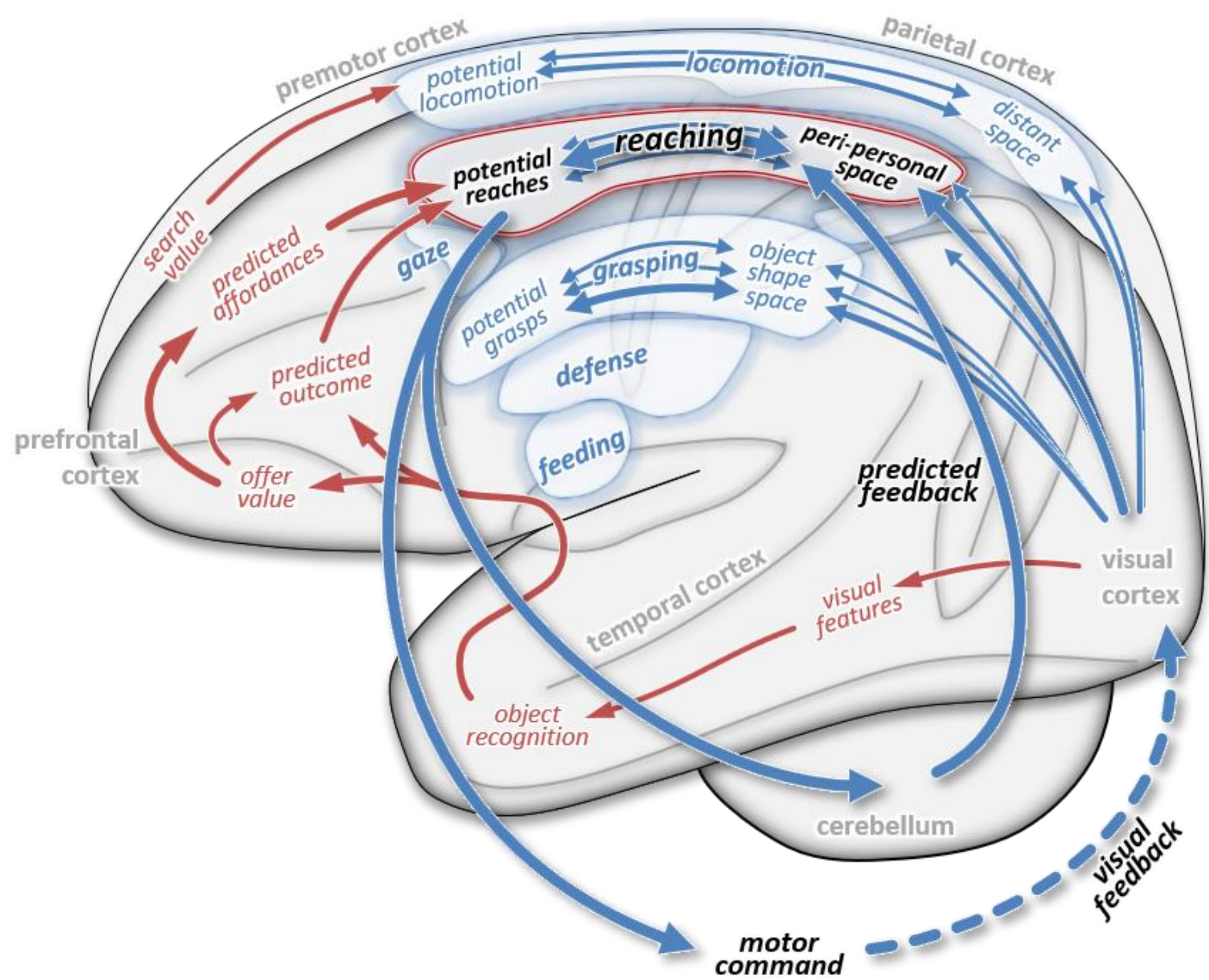

***Figure 3: Schematic of hierarchical affordance competition in the context of visually guided primate behavior.*** *Figure adapted from* [48,49,53]. *Blue lines indicate spatial visual information specifying multiple potential actions in different frontoparietal circuits* [54] *dedicated to different kinds of actions (locomotion, reaching, grasping, etc.). Red lines indicate information used for selecting an action from among all of these options. In this example, the red polygon indicates that the reaching system has been selectively invigorated, and that a competition between potential reaching actions unfolds to select one for overt execution (thick blue lines). This produces a motor command that results in visual feedback, as well as predicted feedback through the cerebellum, guiding the hand toward the target. At the same time, the hand shape is adjusted to the target object shape and the grasping movement selected once the target is reached. Other action possibilities may also be specified (e.g. locomotion in medial regions) but not released into execution.*

Importantly, this circuit extends beyond present affordances to include future ones, enabling the prediction and planning of upcoming action possibilities and the selection between immediately available affordances and those that can be predicted to emerge later [49]. This allows sequential decision-making to be understood as navigation through a landscape of present and future affordances, including both predicted affordances and those that can be actively created through behavior. As an agent moves, explores, and receives feedback, its predictive estimate of available action possibilities becomes increasingly precise, while new affordances may emerge and others disappear. For example, as a lion approaches a gazelle, it may transition from non-reachable to reachable, while the gazelle correspondingly loses escape affordances depending on distance and speed. Goal-directed behavior can then be constructed on top of this affordance-based predictive model, by selecting actions that create or eliminate affordances. For instance, a boxer may move closer to an opponent to enable "jabbability" or a monkey may walk out on a tree branch to make a piece of fruit "acquirable".

A closely related idea is that of *motor simulation*: internal models primarily engaged in action control can be reused in offline mode – temporarily "detached" from the action–perception loop – to support imagination, as well as to vicariously simulate others' actions in order to infer their goals and intentions [19]. This perspective highlights that predictions generated by motor control and prediction systems provide a foundational stepping stone for understanding of the physical and social environment and for what is often described as a grounded theory of other minds. Similarly, neural systems originally involved in sequencing sensorimotor actions, such as Broca's area, also play key roles in language, communication, and music, all of which share hierarchical, syntax-like structure. This convergence highlights the continuity between sensorimotor processing and higher cognitive functions, suggesting that advanced linguistic abilities may repurpose neural mechanisms originally evolved for action organization [58].

***Active perception, exploratory learning, and neuromodulatory control***

Living organisms acquire experience through active interaction with the environment, including the social environment, rather than through passive exposure to externally supplied data. The resulting experiences depend directly on the actions selected, as well as on the organism's motivation to explore and learn [59]. This intrinsic drive to acquire knowledge about the world – even beyond immediate instrumental needs – has been described in multiple ways in AI and cognitive science, including curiosity, intrinsic motivation, epistemic action, uncertainty reduction, information seeking, and open-ended exploration.

The regulation of this exploratory capacity relies on intrinsic systems for motivation, arousal, and information seeking. This can be understood as a "meta" world model, which does not describe the environment per se but rather estimates its expected informativeness, captured by *expected information gain* as a signal of uncertainty reduction and learning progress, and that regulates behavior along a continuum from exploratory to exploitative as appropriate [60,61]. Maintaining an explicit estimate of the model's own epistemic uncertainty and knowledge gaps is also essential for endowing generative AI systems with the capacity to actively seek informative experiences and calibrate confidence in their predictions.

Neuromodulatory systems, in particular the noradrenergic system, play a central role in this regulation by shaping behavioral modes associated with exploration [62]. In such states, arousal is increased, sensory gain is enhanced, and learning rates are elevated, collectively supporting an information-seeking mode in which the organism actively engages with environments rich in learnable structure. At the other extreme lie complementary modes characterized by reduced exploratory drive, such as states associated with behavioral disengagement or learned helplessness, in which interaction and information seeking are suppressed. More broadly, these modes – or "moods" – serve as fundamental regulators of cognition and learning in biological organisms, dynamically balancing exploration and exploitation according to internal estimates of key environmental properties, such as expected informativeness, safety, and risk, computed by higher-order ("meta") world models.

***Allostatic control, emotion, and the interoceptive system***

A central feature of living organisms is their autonomy: they must actively ensure their continued existence by regulating energy and evaluating what is beneficial or harmful for survival. To this end, the brains of advanced animals engage in allostatic (anticipatory) regulation of the body's internal milieu, predicting future needs and preparing for changes before they occur [63]. This capability is supported by the allostatic–interoceptive neural system, which sits at the top of the brain's functional hierarchy and strongly shapes cognition and affect, reflecting the tight coupling between metabolic regulation and cognitive processing [64,65]. In fact, interacting with the world can be seen as the extension of metabolic control through an animal's environment [66–68], taking advantage of available affordances to predictively adjust one's state in the world.

The allostatic–interoceptive system can be understood as a form of world model in which the "world" corresponds to the organism's internal milieu (as well as its external environment). It supports interoceptive inference and control by estimating latent somatic and need-related states underlying interoceptive signals and their relation to exteroceptive states. These include deviations from homeostatic setpoints leading to hunger, thirst, heat, cold, fatigue, and pain, as well as more abstract factors such as the need for shelter, or one's place in a social hierarchy. Importantly, these latent variables are not merely read out from sensory signals; they are *inferred* states that explain interoceptive inputs. In addition to inference, the system supports *control* of these latent somatic variables through a hierarchy of strategies, ranging from low-level autonomic reflexes (e.g., thermoregulation through sweating) to high-level goal-directed behavior (e.g., seeking shelter or purchasing an umbrella). In this way, interoceptive inference and control are tightly coupled [69–71].

Higher-level constructs, such as a persistent sense of self, may emerge from interoceptive world models that continuously infer and regulate somatic states in the service of energy balance and survival [72]. Affective and emotional experiences—such as joy, sadness, or hope—may correspond to inferred interoceptive states grounded in these internal signals, forming the basis of our valenced understanding of the world and shaping the emotional significance we assign to autobiographical memories and imagined future events. Many aspects of social cognition, including understanding others' emotions, empathy, and emotional contagion, are likewise rooted in this system. Motivation may arise from tracking changes in interoceptive variables over time, while dysfunctions such as depression may reflect impaired regulation of these variables, leading to increased allostatic load and inefficient energy management [73–75].

Finally, the allostatic–interoceptive system also serves as a core valuation mechanism. Beyond predicting how the external world evolves, organisms must evaluate whether outcomes are beneficial or detrimental to their internal state. This system provides intrinsic signals of success and failure – good or bad – that arise from the organism's own dynamics rather than externally imposed reward functions. This function of the allostatic–interoceptive system highlights that living organisms are intrinsically autonomous and purposive: needs and goals are present from the outset, rather than being imposed later, as is typically the case in current AI systems [11].

### ***Self–other distinction, corollary discharge, and sensory cancellation***

Another important aspect of autonomy is identifying the consequences of one's own actions and distinguishing self-generated sensory effects from the external dynamics of the environment. Many living organisms, including relatively simple ones, possess neural circuits that serve this function and can be understood as mechanisms for establishing the boundary between "self" and "world" to support more efficient perception within a grounded world model [76–78].

The core principle is that when an action is performed, a copy of the motor command – or of its predicted sensory consequences, known as *efference copy* or *corollary discharge* – is sent to sensory systems. This signal enables the brain to attenuate or cancel predictable self-generated sensory inputs, thereby filtering out the consequences of its own actions and emphasizing information that is more likely to reflect external events.

Beyond shaping perception, these mechanisms may contribute to the construction of higher-order cognitive states such as agency, intentionality, and the sense of control, all of which require inferring whether a particular outcome was caused by oneself or by external factors. Conversely, disruptions of these processes may contribute to disorders involving misattributions of agency, such as failing to recognize one's causal role in an outcome or erroneously attributing external events to oneself [79].

**Discussion**

In current embodied AI, world models have emerged as a fundamental ingredient. The dominant approach is to extend large language model architectures with visual, proprioceptive, and other sensory modalities, so that the linguistic training provides the primary scaffold for world modeling, knowledge organization and reasoning. Biological systems, however, are organized in the opposite way, with world models emerging from sensorimotor interaction to provide the scaffold that gives words their meaning. Here, we highlighted several neural and computational principles underlying the grounded world models of living organisms, which support navigation in physical and conceptual spaces, affordance-based interaction with objects, exploratory learning, allostatic regulation, and self–other distinction.

Several recurrent themes emerge from these examples, including the role of intrinsic dynamics as a foundation for learning, the centrality of action in aligning these dynamics with the external world, the prominence of autonomous experience and open-ended learning over passive assimilation of externally provided data, and the way in which early predictive and control mechanisms scaffold higher cognitive abilities such as reasoning, conceptual navigation, planning, and imagination. Together, these examples illustrate how grounded world models emerge from embodied, autonomous, and continual interaction with the environment through multimodal predictive and control circuits spanning exteroceptive, interoceptive, and proprioceptive domains, and how the progressive reuse and elaboration of these circuits provides the foundation for higher cognition, social behavior, and language. Summing up, these observations portray living organisms as autonomous, active learners of grounded world models, with intrinsic neural dynamics providing effective inductive priors, and sensorimotor prediction and control circuits furnishing the foundation upon which advanced cognitive abilities, including language, are built.

None of these biological insights are currently central to the development of generative or embodied AI, and one may reasonably ask whether they are essential or merely incidental. At present, it is unclear whether the predominantly passive paradigm will eventually encounter fundamental limitations, particularly given the rapid pace at which the field continues to develop new methods and architectures. It is also possible that active and passive learning approaches converge toward similar representations of certain domains, such as space, despite relying on very different learning processes. From an engineering perspective, once the fundamental principles are identified, many biological details may prove unnecessary. The challenge is that we do not yet know with confidence which aspects of biological world-model acquisition are fundamental and which are contingent products of evolution. For this reason, it is useful to consider the potential advantages and limitations of the different routes toward acquiring world models.

A fundamental advantage of current methodologies for developing embodied AI systems and world models, such as visuo-language-action models, is that they are relatively well-defined from an engineering perspective, even if computationally expensive and data hungry. By contrast, neurobiology does not yet offer a sufficiently precise or complete methodological framework to directly constrain engineering design, although recent initiatives to map the brains of living organisms at scale [7] or to probe in greater detail the development of cognitive capabilities such as language [80] may accelerate progress. Moreover, some techniques that are highly effective in modern generative

AI – such as large-scale batch training, backpropagation-based optimization, and GPU-driven parallel computation – may not straightforwardly translate to the biological regime.

At the same time, the biological route has the advantage that phylogenetic development, which grounds higher cognition in more basic embodied processes, leaves no conceptual gaps and ensures that meaning emerges from situated experience [2,81]. A common motivation for introducing concepts such as world models or spatial intelligence is the problem of grounding language in the physical world – for example, enabling a robot to correctly interpret instructions such as "be careful with this fragile cup" and translate them into actions that avoid damaging the object. In contrast, living organisms do not face this problem in the same way, as linguistic meaning is already grounded in sensorimotor experience. Concepts emerge through interaction before being labeled with words, such that terms like "fragile" are intrinsically linked to prior experience manipulating objects in the world.

The biological route might be particularly relevant when developing embodied AI systems capable of engaging in rich and smooth social interaction with humans and aligning with human values and perspectives. This raises the question of whether such capacities can be added late in development or must instead be grounded from the outset by aligning aspects of their core architecture more closely with that of humans. As discussed above, in humans, the ability to represent and act on social motivations such as care for others, or to reliably adhere to ethical constraints, is grounded in affective and interoceptive systems supporting valuation, empathy, and attachment, reflecting our intrinsically social nature [82]. Furthermore, the ability to engage in smooth and legible social interaction depends not only on fluent language but also on action mirroring circuits that align the actions and goals of interacting individuals [83–85] as well as subtle forms of non-verbal communication, including behavioral alignment and the capacity to make oneself predictable and legible to others [21]. Endowing future embodied AI agents with analogous mechanisms – perhaps as constraints or inductive biases – may pave the way for more effective and natural interaction with humans.

Yet another complementary aspect is the possibility of enabling autonomous learning in embodied AI agents within rich social interaction contexts [86]. As a highly social species, humans acquire much of their knowledge and many of their skills not only through individual experience [59] but also through interaction and joint learning. Crucially, social practices enable the construction of *shared world models*, grounded in common representations of language, physical dynamics, object persistence, and culturally embedded values, moral and social norms [86,87]. Even subjective experiences such as emotions are shaped not only by private interoceptive signals but also by social interpretation and sharing [82]. This raises the possibility that embodied AI systems trained within social environments – through continuous interaction with humans and other agents – could develop more naturally aligned representations embedded in human world models and values, potentially enabling safer, more reliable communication and social competence.

Pursuing these research directions would pose significant novel challenges to the field. Current AI research tends to progress fastest along directions that fit the available hardware and software stack. The prominence of language-centric models may partly reflect this bias, since text is discrete, compressed, abundant, and computationally efficient to scale, whereas embodied sensorimotor experience is continuous, temporally dense, interactive, and expensive to process. Rather, training embodied AI systems under biologically inspired and socially grounded learning regimes will likely require substantial algorithmic advances, particularly in active learning, self-generated curricula, continual learning, and interactive learning. At the same time, introducing objectives and inductive biases that favor compact, action-relevant, grounded world models could help explore forms of intelligence that are currently underrepresented by the prevailing hardware lottery, moving beyond a conception of machine intelligence shaped primarily by contemporary computational constraints [88].

Summing up, the biological perspective on grounded world models highlights how cognition, action, emotion, and social understanding emerge from predictive architectures that autonomously engage with the physical and social environment to pursue pressing goals. This contrasts with current generative and embodied AI approaches, which remain largely rooted in passive, language-centric training paradigms and largely lack autonomy and goals. While it remains unclear which aspects of biological organization are essential versus incidental for building intelligent systems, these considerations suggest that future progress may benefit from integrating principles of predictive processing, autonomous learning, and biological regulation alongside existing engineering approaches.

**Acknowledgements**

This research received funding from the European Research Council under the Grant Agreement No. 820213 (ThinkAhead), the Italian National Recovery and Resilience Plan (NRRP), M4C2, funded by the European Union – NextGenerationEU (Project IR0000011, CUP B51E22000150006, “EBRAINS-Italy”; Project PE0000013, “FAIR”; Project PE0000006, “MNESYS”), and the PRIN PNRR P20224FESY. We used a Generative AI model to correct typographical errors and edit language for clarity.